\documentclass{article}
\usepackage[english]{babel}
\usepackage{float}
\usepackage{booktabs}
\usepackage{graphicx}
\usepackage{dcolumn}
\usepackage{bm}
\usepackage{cite}
\usepackage{amsthm} 
\include{cip-v3-pack-and-defined}

\usepackage[letterpaper,top=2cm,bottom=2cm,left=3cm,right=3cm,marginparwidth=1.75cm]{geometry}

\usepackage{amsmath}
\usepackage{graphicx}
\usepackage[colorlinks=true, allcolors=blue]{hyperref}
\usepackage{authblk}
\usepackage{multirow}

\title{Quantifying discriminability of evaluation metrics in link prediction for real networks}
\author[a,1]{Shuyan Wan}
\author[a,1]{Yilin Bi}
\author[a]{Xinshan Jiao}
\author[a,$\ast$]{Tao Zhou}
\affil[a]{CompleX Lab, School of Computer Science and Engineering, University of Electronic Science and Technology of China, Chengdu 610054, China}
\affil[1]{Shuyan Wan and Yilin Bi contributed equally to this work.}
\affil[$\ast$]{To whom correspondence should be addressed: \href{zhutou@ustc.edu}{zhutou@ustc.edu}}

\begin{document}
\maketitle

\begin{abstract}
Link prediction is one of the most productive branches in network science, aiming to predict links that would have existed but have not yet been observed, or links that will appear during the evolution of the network. Over nearly two decades, the field of link prediction has amassed a substantial body of research, encompassing a plethora of algorithms and diverse applications. For any algorithm, one or more evaluation metrics are required to assess its performance. Because using different evaluation metrics can provide different assessments of the algorithm performance, how to select appropriate evaluation metrics is a fundamental issue in link prediction. To address this issue, we propose a novel measure that quantifiers the discriminability of any evaluation metric given a real network and an algorithm. Based on 131 real networks and 20 representative algorithms, we systematically compare the discriminabilities of eight evaluation metrics, and demonstrate that H-measure and Area Under the ROC Curve (AUC) exhibit the strongest discriminabilities, followed by Normalized Discounted Cumulative Gain (NDCG). Our finding is robust for networks in different domains and algorithms of different types. This study provides insights into the selection of evaluation metrics, which may further contribute to standardizing the evaluating process of link prediction algorithms.
\end{abstract}


\section{Introduction}
Network is a powerful mathematical tool to characterise complex interactions among multiple entities, ranging from natural systems to human societies. In a network, nodes represent entities (such as individuals, companies, genes, etc.), and links represent interactions between these entities (such as friendships, trading partnerships, gene regulatory relationships, etc.). Networks are ubiquitous, spanning from interpersonal interactions in daily life to international commodity exchanges, and from molecular interactions within cells to information transfer through the Internet. Against this backdrop, network science has emerged to study the structure, evolution and dynamics of networks from diverse fields, with the aim of developing universal theories, models, algorithms and tools to uncover common properties among seemingly disparate networks \cite{barabasi2016,newman2018}.

Link prediction is one of the most productive branches in network science. It aims to predict missing or future links by analyzing the network structure \cite{liu2011,wang2015,Martínez2016,kumar2020,divakaran2020,zhou2021a,arrar2024}. Link prediction has found a wide range of theoretical and practical applications, such as quantitatively evaluating the accuracy of network evolution models \cite{Wang2012,Zhang2015}, personalized recommendation of friends and products \cite{aiello2012,liu2012}, identifying spam emails \cite{lin2003,rattigan2005}, predicting the pathways and speeds of disease spreading \cite{akhtar2023}, detecting high-risk transactions in the financial sector \cite{wang2024}, predicting potential interactions between proteins or genes \cite{csermely2013,ding2014}, forecasting traffic flow distribution \cite{lu2020}, and discovering potential opportunities for scientific collaboration \cite{lande2020}, among others.

The past decade has witnessed the emergence of a large number of link prediction algorithms, with some representative examples as Refs.\cite{hasan2006,libennowell2007,clauset2008,guimerà2009,zhou2009,liu2010,menon2011,cannistraci2013,liu2015,pan2016,pech2017,zhang2018,benson2018,kovács2019,kitsak2020}. Consequently, how to select appropriate evaluation metrics to fairly assess the performance of these algorithms, and thereby correctly guide the direction of algorithm design and optimization, has become a critical issue in the field of link prediction \cite{lichtenwalter2010,yang2015,zhou2021a}. Different evaluation metrics typically possess distinct attributes and are suitable for specific scenarios. For example, Precision focuses on the accuracy of positive predictions and is particularly applicable in scenarios where the cost of false positives (predicting as positive when actually negative) is high, such as in medical diagnostics and financial risk assessment \cite{buckland1994}. Matthews Correlation Coefficient (MCC) is suitable for situations where the data categories are imbalanced or the number of samples is not uniform, such as drug screening and credit scoring \cite{matthews1975}. NDCG considers the importance of each position in the ranking of predictions, which is mainly used in ranking tasks such as recommender systems and search engines \cite{Jarvelin2002}. AUC reflects the ability of the model to distinguish between positive and negative samples, and it is effective in evaluating the overall performance of the model \cite{hanley1982}. Area Under the Precision-Recall Curve (AUPR) is suitable for tasks with imbalanced classes, such as anomaly detection \cite{davis2006}. AUC-precision refers to the area under the precision curve for the top-L positions, assessing how effectively positive links are prioritized within the top-L positions \cite{muscoloni2022}.

Among the aforementioned metrics, AUC is the most widely used and has consequently faced criticism from some scholars \cite{lobo2008}. Yang \textit{et al.} argued that in cases of severe class imbalance between positive and negative samples, AUPR is preferable to AUC \cite{yang2015}. Lichtenwalter and Chawla pointed out that AUC tends to overrate algorithms that can rank many negative samples at the bottom, which is not meaningful for identifying the desired positive samples \cite{lichtenwalter2012}. Saito and Rehmsmeier emphasized the limitations of AUC in scenarios where only the top-ranked predictions are valuable (referred to as early retrieval problems, which is common in search and recommendation) \cite{saito2015}; this limitation has also been noted in the evaluation of binary classification algorithms \cite{baker2001}. To address potential shortcomings of AUC, several improved versions have been proposed. For example, Hand pointed out that: AUC uses different misclassification cost matrices when comparing different classifiers. Based on this, Hand proposed a metric called H-measure, which adopts the core concept of AUC while ensures consistent misclassification cost matrices across different classifiers \cite{hand2009}. Muscoloni \textit{et al.} argued that the importance of earlier positions is greater than that of later ones, leading them to propose a new metric called AUC-mROC, based on rescaling the horizontal and vertical coordinates of the ROC curve \cite{muscoloni2022}.

\begin{figure*}[htbp]
	\centering
	\centerline{\includegraphics[width=1\textwidth]{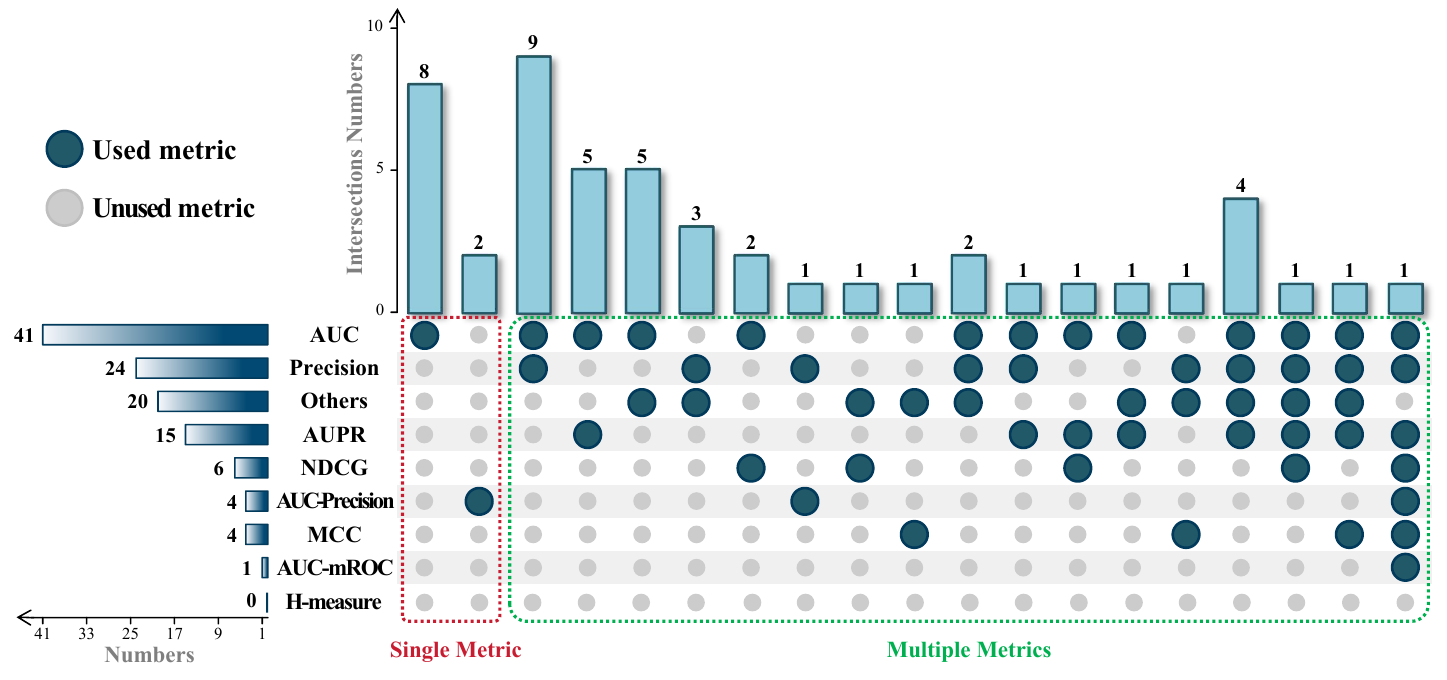}}
	\caption{{\bf The survey of the evaluation metrics used in link prediction studies.} We selected a total of 50 papers on link prediction, including well-known works \cite{hasan2006, libennowell2007, clauset2008, liu2009, zhou2009, guimerà2009, liu2010, menon2011, dunlavy2011, lichtenwalter2012, cannistraci2013, kuo2013, liu2015, sherkat2015, yang2015, wang2015, trouillon2016, pan2016, pech2017, muscoloni2018, zhang2018, ding2018, yasami2018, kastrin2018, kovács2019, chen2019, pech2019, kumar2019, haghani2019} and some selected studies in the past five years \cite{ahmad2020, kitsak2020, ghasemian2019, liu2020, nasiri2021a, nasiri2021b, kou2021, chen2021, daza2021, berahmand2021, fan2022, cai2022, guo2022, peng2022, balogh2022, nasiri2023, yao2023, mueller2023, hong2023, muscoloni2023, menand2024}, and compiled the usage of evaluation metrics in these studies.}
	\label{fig1}
\end{figure*}

Given the multitude of evaluation metrics available, how do researchers actually choose them? As shown in Fig. \ref{fig1}, we survey high-influential papers as well as some selected studies in the last five years, and find considerable variation in the choices of evaluation metrics by researchers. Some metrics are widely used; for example, approximately 82$\%$ of the studies use AUC as one of their evaluation metrics, while at the same time, some metrics go unnoticed, e.g., H-measure has never been used. About 20\% of the studies opted for a single metric, while a few studies used more than five evaluation metrics to assess the algorithm performance (note that in Fig. \ref{fig1}, some studies have more than one metric categorized under \textbf{others}). For instance, Kou \textit{et al.} used six evaluation metrics \cite{kou2021}, and Muscoloni and Cannistraci selected seven evaluation metrics that account for prediction accuracy, ranking performance, and class imbalance \cite{muscoloni2023}. Although to select many metrics can effectively mitigate bias associated with relying on a single metric, it consumes more computational resources, and complicates the judgment of which evaluation metric is more credible in cases of inconsistency among the metrics. For example, aside from the best-performed algorithm proposed by Kou \textit{et al.}, the other compared algorithms exhibited inconsistent performance across the six evaluation metrics used in \cite{kou2021}. In fact, a recent large-scale empirical study has demonstrated significant inconsistencies among different evaluation metrics \cite{bi2024}. That is, an algorithm \textbf{A} might outperform algorithm \textbf{B} according to one evaluation metric, but \textbf{B} might outperform \textbf{A} when using another metric.

In summary, we currently face a situation that we know different evaluation metrics can provide inconsistent assessments of algorithm performance, but we lack effective guidance to select suitable metrics, resulting in significant diversity in real usage, as shown in Fig. \ref{fig1}. This diversity may be resulted from a combination of many factors, such as ad hoc selections, herd selections, and even biased selections favoring ones' own algorithms. As mentioned above, some researchers have analysed the advantages and disadvantages of some evaluation metrics, but there is still a lack of studies that quantitatively characterise the specific abilities of various metrics and perform comparative analysis across different metrics. As the key of the quality of an evaluation metric lies in its ability to correctly distinguish between algorithms with similar performance, we have proposed a method to characterise the discriminabilities of evaluation metrics under a very specific link prediction algorithm for a toy network, where the prediction accuracy is fully controllable \cite{zhou2023,jiao2024}. Although these works provide a starting point, there are still two obvious shortcomings. First, the proposed method cannot be applied to real networks and algorithms. Second, these works have not proposed a single quantitative measure for the discriminability of a metric, making subsequent extension and analysis difficult. In this paper, we propose a new approach that can quantify the discriminability of an arbitrary evaluation metric with a single measure, given real networks and algorithms. Based on 131 real networks from disparate domains and 20 representative algorithms ranging from mechanistic models to machine learning methods, this study systematically compares the discriminabilities of eight evaluation metrics: Precision, MCC, NDCG, AUC, AUPR, AUC-Precision, H-measure, and AUC-mROC. We found that H-measure and AUC exhibit the strongest discriminabilities, followed by NDCG. This work provides a valuable step towards achieving the long-sought consensus in the network science community on how to fairly evaluate link prediction algorithms.

\section{Results}
\subsection{Method Summary}
A graph can be represented as $G(V,E)$, where $V$ denotes the set of nodes, and $E$ denotes the set of links. A link is formed between two nodes if there is a certain relationship or interaction between them. In this paper, we only consider undirected networks, where directions and weights of links are not considered, and multiple connections and self-connection are not allowed. If the network contains $N=|V|$ nodes, then at most $N(N-1)/2$ links exist. The set of all these possible links is denoted as $U$. The task of link prediction is to predict which links in $U-E$ are missing links, i.e., links that exist but have not been observed. To evaluate the performance of link prediction algorithms, we split $E$ into a training set $E^{T}$ and a test set $E^{P}$ (sometimes a part of $E^{T}$ is further sampled out as the validation set to determine model parameters), and ensure that $E=E^{T}\cup E^{P}$, $E^{T}\cap E^{P}=\emptyset$. The links in the training set $E^{T}$ are known, while the links in the test set $E^{P}$ are treated as unknown. Obviously, a good link prediction algorithm tends to accurately find out the test set $E^{P}$ from the set of all unknown links, say $U-E^{T}$.

\begin{figure*}[htbp]
	\centering
\centerline{\includegraphics[width=1\textwidth]{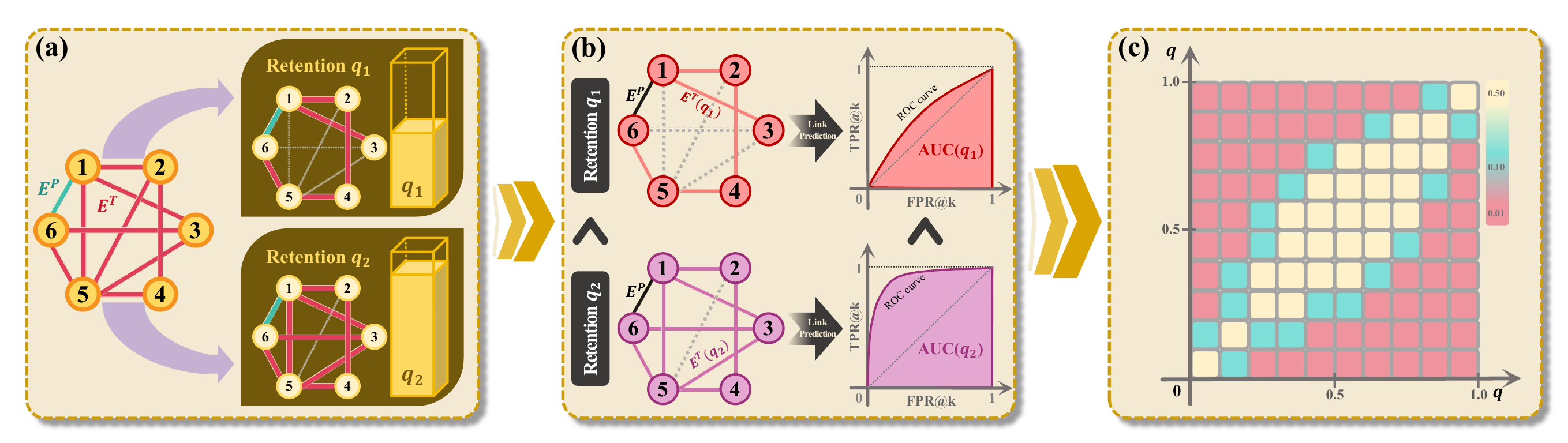}}
	\caption{{\bf The procedure of this study.} (a) shows the extraction of links from the training set $E^{T}$ under different retention rates $q_{1}$ and $q_{2}$ for the same network, where $q_{1} < q_{2}$. (b) illustrates the calculation of evaluation metric (using AUC as an example) under different retention rates $q_{1}$ and $q_{2}$. (c) depicts the process to obtain the $p$-value matrix $[p_{ij}]_{n \times n}$ by comparing scores assigned by the evaluation metric under different retention rates $q$ based on a large number of independent experiments, where the area of the red region is the discriminability $d$ for the evaluation metric.}
	\label{fig2}
\end{figure*}

In this study, for any target network we randomly divide the link set to a training set $E^{T}$ and a test set $E^{P}$, ensuring that the ratio of links in the training set to those in the test set is $|E^{T}|:|E^{P}|=9:1$. The links in $U-E$ are labeled as negative samples for subsequent performance evaluation. A natural assumption is that given a link prediction algorithm and the set of negative samples (i.e., the set of negative samples is kept unchanged), the more links being observed (i.e., richer information is known), the more accurate the prediction will be. Accordingly, we randomly select a proportion $q$ of links in $E^{T}$ for training, while the remaining $(1-q) \cdot |E^{T}|$ links are completely discarded (as shown in Fig. \ref{fig2}(a)), with the parameter $q \in (0,1)$ referred to as the retention rate. Note that, the discarded $(1-q) \cdot |E^{T}|$ links are not added to the negative samples, so that both the positive and negative samples (i.e., $E^{P}$ and $U-E$) remain unchanged, avoiding the uncertainty of the algorithm's performance due to the change in sample structure. Let $M$ denotes the evaluation metric, and $\Omega$ denotes the link prediction algorithm, and $M[\Omega(q)]$ denotes the score of algorithm $\Omega$ under the metric $M$ with the retention rate $q$. Given a target network, if $q_{i}<q_{j}$, a sufficiently discriminative metric $M$ should assign scores that satisfy the inequality $M[\Omega(q_{i})]<M[\Omega(q_{j})]$ (as shown in Fig. \ref{fig2}(b)). If we conduct $T$ independent experiments (randomly reallocating the training and test sets each time) to test the above inequality, and in $t$ of these experiments the results do not satisfy the assumption (i.e., $M[\Omega(q_{i})] \geq M[\Omega(q_{j})]$), then the corresponding $p$-value is $p_{ij}=\frac{t}{T}$. If $p_{ij}$ is lower than a predefined threshold $p^\ast$, it indicates that the evaluation metric $M$ can effectively distinguish between algorithm performance with retention rates $q_{i}$ and $q_{j}$ at the significance level $p^\ast$. If we consider a finite number of retention rates, denoted as $q_{1}<q_{2}<\cdots<q_{n}$, by examining all pairs $(q_{i},q_{j})$ and calculating the corresponding $p_{ij}$ (if $q_{i}>q_{j}$, let $p_{ij}=p_{ji}$), we can obtain a binary discriminability matrix $S^{n\times n}$: 
\begin{equation}
	s_{ij}=
	\begin{cases}
		1, \quad p_{ij} < p^\ast\\
		0, \quad p_{ij} \geq p^\ast
	\end{cases}.
\end{equation}
Based on this, we can define the discriminability of the evaluation metric $M$ for the link prediction algorithm $\Omega$ at the confidence level $p^\ast$ as:
\begin{equation}
	d(M,\Omega,p^\ast)=\frac{\sum_{1\leq i,j \leq n}s_{ij}}{n^2}.
\end{equation}
\label{value}
As shown in Fig. \ref{fig2}(c), this equation actually calculates the proportion of cells in the discriminability matrix $S$ that take the value of 1.

The experiments in this study encompass 131 real networks across six domains, including 29 biological networks, 18 economic networks, 17 information networks, 24 social networks, 25 technological networks, and 18 transportation networks (see \textbf{Table S1 in SI Appendix} for detailed information). We performed a comparative analysis of eight commonly used evaluation metrics: Precision, MCC, NDCG, AUC, AUPR, AUC-Precision, H-measure, and AUC-mROC (detailed definitions are provided in \textbf{Methods}). A total of 20 representative link prediction algorithms, comprising both similarity-based and embedding-based algorithms, were evaluated. These algorithms are further categorized into six subclasses, as detailed in Table \ref{tab1}.

\begin{table}[htbp]
	\renewcommand{\arraystretch}{1.2} 
	\centering
	\caption{\textbf{Link prediction algorithms.}}
	\scalebox{0.7}{%
		\begin{tabular}{cccc}
			\hline
			\multicolumn{2}{c}{\textbf{Classification}}      & \textbf{Algorithms}       & \textbf{Reference}  \\ \hline
			\multirow{13}{*}{Similarity-based} & \multirow{4}{*}{\begin{tabular}[c]{@{}c@{}}Local Similarity\\ Indices\end{tabular}}                                                                & Common Neighbor Index (CN)   & \cite{newman2001}     \\                       
			&   & Resource Allocation Index (RA) & \cite{zhou2009}     \\
			&   & Jaccard Simialrity (JA) & \cite{jaccard1901}    \\
			&   & Preferential Attachment (PA)    & \cite{barabasi2002}   \\ \cline{2-4}
			& \multirow{3}{*}{\begin{tabular}[c]{@{}c@{}}Global Similarity\\ Indices\end{tabular}}      
			& Katz Index (KA)     & \cite{katz1953}       \\
			&   & Matrix Forest Index (MFI)   & \cite{chebotarev1997} \\
			&   & SimRank (SR)     & \cite{jeh2002}        \\ \cline{2-4}
			& \multirow{6}{*}{\begin{tabular}[c]{@{}c@{}}Quasi-local Similarity\\ Indices\end{tabular}} & \begin{tabular}[c]{@{}c@{}}L2-based Cannistraci-Hebb network\\ automaton model number two (CH2-L2)\end{tabular} & \cite{muscoloni2018}  \\
			&   & L3-based Common Neighbor (CN-L3)    & \cite{zhou2021b}      \\
			&   & L3-based Resource Allocation (RA-L3)     & \cite{zhou2021b}      \\
			&   & \begin{tabular}[c]{@{}c@{}}L3-based Cannistraci-Hebb network\\ automaton model number two (CH2-L3)\end{tabular} & \cite{muscoloni2018}  \\
			&   & Local Random Walk (LRW)     & \cite{liu2010}        \\
			&   & Superposed Random Walk (SRW)      & \cite{liu2010}        \\ \hline
			\multirow{7}{*}{Embedding-based}   & Matrix Factorization    & NetMF (NMF)     & \cite{qiu2018}        \\ \cline{2-4}
			& \multirow{2}{*}{Random Walk}     & DeepWalk (DW)    & \cite{perozzi2014}    \\
			&   & Node2Vec (N2V)      & \cite{grover2016}     \\ \cline{2-4}
			& \multirow{4}{*}{Deep Learning}       
			& Graph Convolutional Networks (GCN)   & \cite{kipf2017}       \\
			&   & Graph Attention Networks (GAT)    & \cite{velickovic2018} \\
			&   & GraphSAGE (SAGE)      & \cite{hamilton2017}   \\
			&   & Variational Graph Normalized Auto-Encoders (VGNAE)    & \cite{ahn2021}        \\ \hline
			\label{tab1}
		\end{tabular}
  }
\end{table}

\subsection{Comparing Discriminabilities of Evaluation Metrics}
\begin{figure}[htbp]
	\centering
	\centerline{\includegraphics[width=0.65\linewidth]{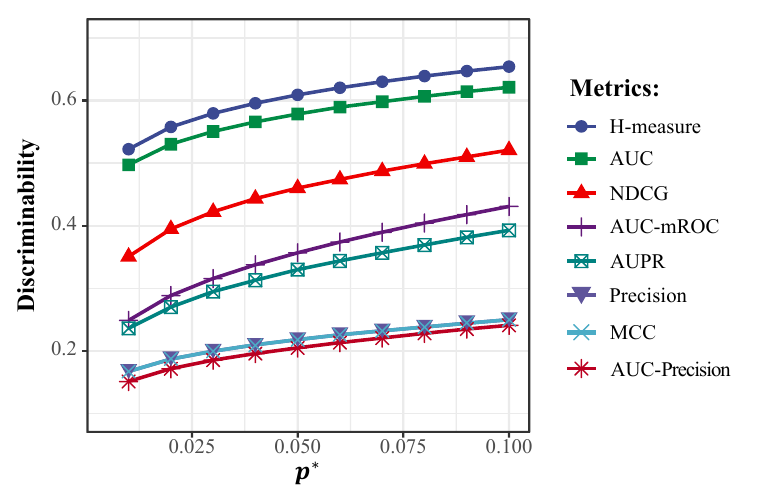}}
	\caption{{\bf The average discriminabilities of evaluation metrics with the varying threshold $p^\ast$.} The results are obtained by averaging over 20 link prediction algorithms and 131 real networks. Here, we set $T=100$.}
	\label{fig3}
\end{figure}

Some recent large-scale empirical studies have demonstrated significant variations in algorithm performance across networks from different domains \cite{zhou2021b,muscoloni2023}. Consequently, to enhance the applicability of the findings of this study, we selected 131 real networks from six distinct domains. Figure \ref{fig3} illustrates how the average discriminability of each metric varies with $p^\ast$ over 131 real networks and 20 algorithms. As depicted in Fig. \ref{fig3}, the discriminabilities of the eight evaluation metrics can be broadly categorized into four tiers. H-measure and AUC have the strongest discriminabilities, followed by NDCG. The third tier includes AUC-mROC and AUPR, while AUC-Precision, Precision and MCC have the weakest discriminabilities. Based on these findings, when confronted with the task of link prediction, we recommend prioritizing H-measure, AUC, and NDCG. Conversely, caution should be exercised when using AUC-Precision, Precision and MCC.

\begin{figure}[htbp]
	\centering
	\centerline{\includegraphics[width=1\linewidth]{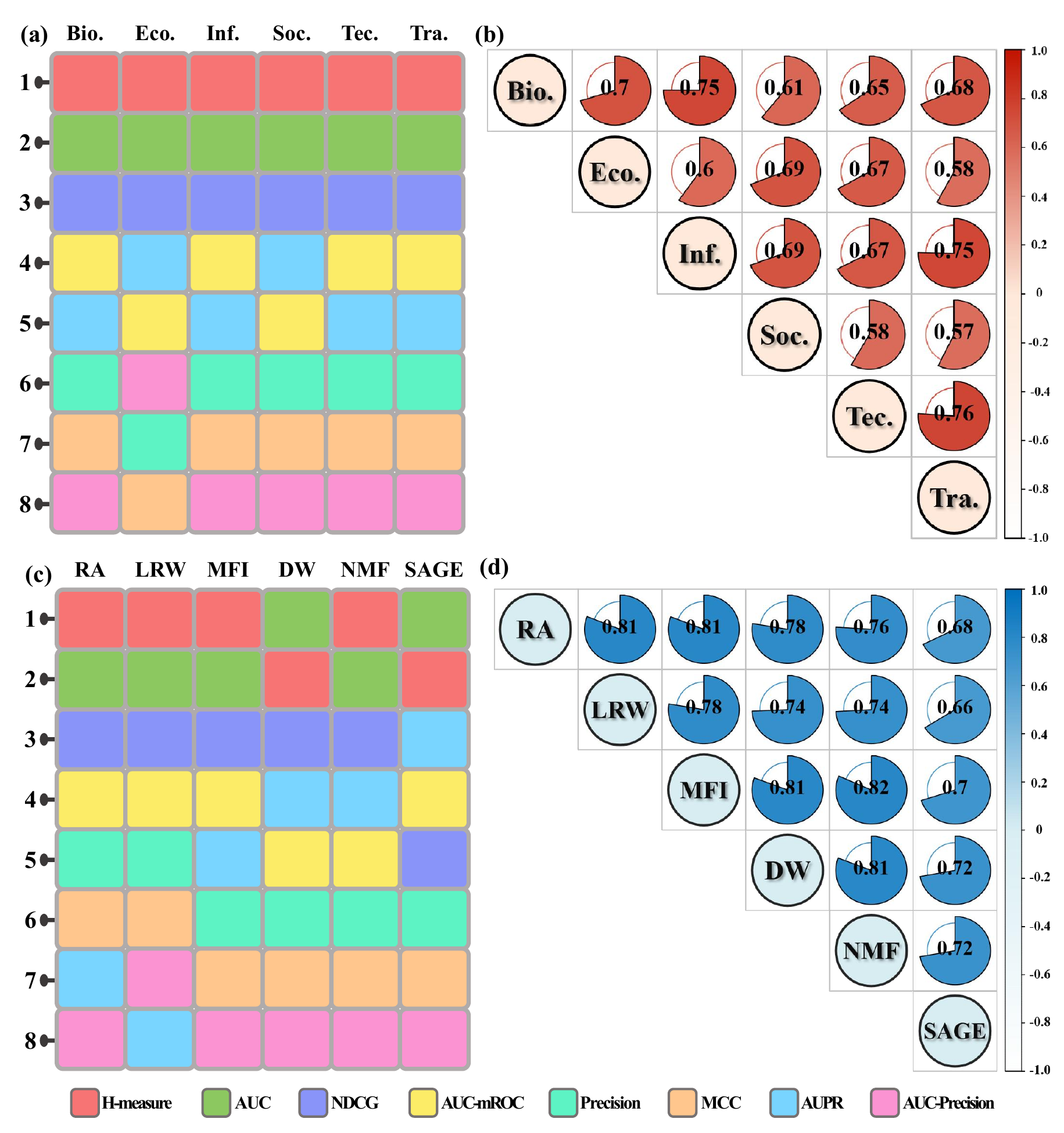}}
	\caption{{\bf The rankings and correlations of discriminabilities of evaluation metrics across different domains and link prediction algorithms.} In (a) and (d), each column represents one domain or one algorithm, with the numbers on the left indicating the ranks of the discriminabilities within the column. (a) shows the rankings of average discriminabilities for the six domains over the 20 considered algorithms, where Bio., Eco., Inf., Soc., Tec. and Tra. represent the domains of biology, economics, information, sociology, technology and transportation, respectively. (d) displays the rankings of the average discriminabilities for six selected algorithms over the 131 real networks. (b) illustrates the gray correlation coefficients for any two domains using the RA index. (d) shows the gray correlation coefficients for all algorithm pairs from the six selected algorithms. In both cases, darker colors indicate higher correlations. Here, we set $p^{\ast}=0.01$ and $T=100$.}
	\label{fig4}
\end{figure}

\subsection{Domain-Specific Analysis}
A link prediction algorithm may exhibit unique applicability for networks in a certain domain. For instance, link prediction algorithms that perform well in social networks often focus on capturing social interaction patterns among users \cite{daud2020}, such as common friends, group affiliations, and communication frequency, etc. In contrast, algorithms that excel in biological networks tend to focus on functional relationships and mechanisms within organisms \cite{teji2023}, exploiting co-expression, functional similarities and other bioinformatics features. This phenomenon prompts us to consider a similar question: Does the domain-specific nature of networks affect the discriminabilities of evaluation metrics? Could a certain evaluation metric demonstrate strong discriminability for networks in one domain while weak discriminability for networks in another domain? Figure \ref{fig4}(a) illustrates the rankings of average discriminabilities for networks across six different domains, where it can be seen that H-measure, AUC and NDCG consistently rank in the top across different domains. Notably, in biological, information, technological, and transportation networks, the rankings of the discriminabilities of the eight metrics are consistent, indicating that our assessment method for discriminability has good generalizability across different domains (see \textbf{Fig. S1 in SI Appendix}, where we present a detailed illustration of the distributions of discriminabilities of the eight evaluation metrics in different domains). Furthermore, we have calculated the correlation coefficients between all domain pairs (see \textbf{Methods} about the grey correlation coefficient \cite{deng1982} applied in this study, which is particularly good at measuring correlation between two sequences with very few elements). As shown in Fig. \ref{fig4}(a), the sequences of metrics' discriminabilities for different domains are highly correlated to each other, again indicating the robustness of our finding.

\subsection{Algorithm-Specific Analysis}
In the previous two subsections, we observed that H-measure, AUC and NDCG possess the strongest discriminabilities and consistently rank among the top three across various network domains. This subsection further examines the robustness of the primary conclusion by checking whether these three metrics again perform the best across different types of algorithms. According to the categorisation of the algorithms in Table \ref{tab1}, we select one representative algorithm from each category (i.e., RA, LRW, MFI, DW, NMF and SAGE) and obtain the rankings of discriminabilities of the eight metrics under these algorithms. Figure \ref{fig4}(c) presents the rankings of the average discriminabilities of the eight evaluation metrics over 131 real networks for each of these six representative algorithms. As shown in Fig. \ref{fig4}(c), H-measure and AUC are consistently ranked as the top two evaluation metrics for every considered algorithm. Specifically, H-measure ranks first for RA, LRW, MFI and NMF, whereas AUC ranks first for DW and SAGE. Additionally, NDCG ranks third in five out of the six algorithms, with its overall performance significantly surpassing the other five evaluation metrics. Similar to Fig. \ref{fig4}(b), we show the correlations between the sequences of metrics' discriminabilities for different algorithms in Fig. \ref{fig4}(d), which are even higher than those correlations between different domains. In a word, these results indicate that our primary finding is robust across different types of algorithms (see also \textbf{Fig. S2 in SI Appendix}, where we provide a detailed illustration of the distributions of the discriminabilities of the eight evaluation metrics for different link prediction algorithms).

\section{Discussion}
In this paper, we propose a method to quantitatively measure the discriminability of evaluation metrics for real networks and practical algorithms, and conduct large-scale experiments that demonstrate that H-measure, AUC and NDCG as the top-3 metrics subject to the discriminability. AUC, which ranked second, is the most widely used yet contentious metric in the field of link prediction, and the present finding can bolster the confidence in its applications. Although H-measure has made significant improvements to AUC \cite{hand2009}, it has not got enough attention commensurate with its scientific merit \cite{xia2017,zian2021,hand2023}, and there are virtually no direct applications of H-measure in link prediction. This paper verifies the excellent performance of H-measure from the perspective of discriminability, so we hope that researchers will pay more attention to H-measure and consider it as an alternative evaluation metric when assessing link prediction algorithms. NDCG has found limited applications in link prediction and related problems \cite{ lichtenwalter2010,yang2012}, and it is worth increasing attention in the future.

We have verified that the primary finding remains valid across different domains of networks and various types of algorithms. For embedding algorithms, we further examine the impact of the key parameter, the embedding dimension, on the discriminability. As illustrated in \textbf{Fig. S3 of SI Appendix}, the relative ranking of the metric discriminabilities is insensitive to the embedding dimension, again validating the robustness of the main conclusion. We also analyze the discriminability of each evaluation metric in synthetic networks \cite{zhou2023,jiao2024}. As depicted in \textbf{Fig. S4 of SI Appendix}, H-measure, AUC, NDCG and AUPR are in the highest discriminability tier. Although this result exhibits minor discrepancies compared to real networks, the overall consistency again supports our main finding.

In summary, this paper quantitatively analyzed the discriminability of several well-known evaluation metrics. The result offers valuable insights into reaching a consensus on how to choose appropriate evaluation metrics to assess algorithm performance and could potentially contribute to the development of related guidelines and standards. Furthermore, the method proposed in this study can be easily extended to various classification problems.

\section{Methods}
\subsection{Evaluation Metrics}
Links in $U-E^{T}$ can be ranked according to their assigned scores, where a higher score indicates a higher existing likelihood. Without loss of generality, the top-$L$ links are predicted to be positive (i.e., missing links), while the remaining links are predicted as negative (i.e., links that are considered not to exist). By comparing these predictions with the ground truth, a confusion matrix can be constructed: within $E^{P}$, the number of links predicted correctly is classified as True Positive (TP), while the number of links predicted incorrectly corresponds to False Negative (FN). In contrast, within $U-E$, True Negative (TN) and False Positive (FP) represent the correctly predicted negative links and incorrectly predicted positive links.

Precision is the ratio of correctly predicted positives to the total number of predicted positives \cite{buckland1994}. In link prediction, the algorithm ranks all links by their existing likelihoods, and the top-$L$  ranked links are considered to be positives (i.e., missing links). Precision is then calculated as the proportion of actual missing links among these top-$L$  predicted links, as:

\begin{equation}
	Precision=\frac{TP}{TP+FP}=\frac{TP}{L}, 
	\label{Precision@k}
\end{equation}
where $TP$ and $FP$ respectively refer to the number of missing links and non-existent links among the top-$L$ predicted links.

MCC (Matthews Correlation Coefficient) measures the correlation between actual outcomes and predicted results \cite{matthews1975}. It is particularly useful for evaluating algorithm performance for imbalanced learning. MCC can be calculated as:

\begin{equation}
	MCC=\frac{TP\cdot TN-FP \cdot FN }{\sqrt{(TP + FP)\cdot (TP + FN) \cdot (TN + FP)\cdot (TN + FN)} }. 
	\label{MCC}
\end{equation}
MCC ranges from -1 to 1, where 1 indicates perfect prediction, 0 reflects random guessing, and -1 represents complete misclassification. The metric is symmetric, meaning it remains unchanged even if positive and negative labels are swapped.

For Precision and MCC, the threshold value $L$ is set $|E^{P}|$.

NDCG (Normalized Discounted Cumulative Gain) evaluates the quality of rankings by giving higher scores to missing links ranked higher \cite{Jarvelin2002}. It uses a logarithmic discounting mechanism, as:

\begin{equation}
	NDCG=\sum_{i=1}^{\left | E^P  \right | } \frac{1}{\log_{2}{(1+r_i)}} 
	\bigg/ \sum_{r=1}^{\left | E^P  \right | } \frac{1}{\log_{2}{(1+r)}},
	\label{NDCG}
\end{equation}
where $r_i$ represents the rank of the $i$-th missing link. 

AUC (Area Under the ROC Curve) represents the probability that a randomly chosen positive sample (missing link) is ranked higher than a randomly chosen negative sample (non-existent link).  In a sorted sequence of $|U-E^T|$ potential links, let the positions of the $|E^P|$ missing links be $r_1 < r_2 < \dots < r_{|E^P|}$. The probability of the $i$-th missing link ranking higher than negative samples is $1 - (r_i - i) / |U-E|$. AUC is calculated by averaging these probabilities:

\begin{equation}
	AUC=\frac{1}{\left | E^{P} \right | }\sum_{i=1}^{\left | E^P \right | } \left ( 1-\frac{r_{i}-i}{\left | U-E \right | } \right ) =1-\frac{\left \langle r \right \rangle }{\left | U-E \right | }+\frac{\left | E^{P} +1\right | }{2\left | U-E \right | },
\end{equation}
where $\left \langle r \right \rangle$ is the average rank of the missing links. In sparse networks, where $|E^P| / |U-E| \ll 1$, this can be approximated as:

\begin{equation}
	AUC \approx 1-\frac{\left \langle r \right \rangle }{\left | U-E^{T} \right | },
	\label{AUC}
\end{equation}
which is equivalent to the so-called ranking score \cite{AUC2007}.\\

AUPR is the area under the Precision-Recall (PR) curve \cite{davis2006}. The PR curve consists of the points $(Recall@k, Precision@k)$, where $k \in [1,|E^{P}|]$, and Recall is the proportion of links in $E^{P}$ that are predicted as positive samples. AUPR can be calculated by the following equation:
\begin{equation}
	AUPR=\frac{1}{2 \cdot |E^{P}|}(\sum\limits_{i=1}^{|E^{P}|}\frac{i}{r_i}+\sum\limits_{i=1}^{|E^{P}|}\frac{i}{r_{i+1}-1}).
\end{equation}

AUC-Precision calculates the area under the curve composed of the points $(k, Precision@k)$, where $k \in [1,|E^{P}|]$, as \cite{muscoloni2022}:
\begin{equation}
	AUC-Precision=\frac{1}{|E^{P}|-1}\sum\limits_{k=1}^{|E^{P}|}Precision@k=\frac{1}{|E^{P}|-1}\sum\limits_{k=1}^{|E^{P}|}\frac{TP@k}{k}.
\end{equation}

H-measure accounts for the cost of misclassification. Let $\nu_{k}$ represent the cost of misclassifying the $k-$th class, and $\nu=\frac{\nu_{0}}{\nu_{0}+\nu_{1}}$ denote the ratio of the cost of misclassifying a positive sample to that of a negative sample. H-measure employs a Beta distribution for the cost ratio, ensuring it is independent of the score distribution. This approach also preserves the objectivity of the metric, meaning that different researchers will obtain consistent results when using the same dataset. H-measure is defined as follows:
\begin{equation}
	H-measure=1-\frac{L_{\alpha,\beta}}{L_{Max}}=1-\frac{\int Q(T(\nu);\mu,\nu)u_{\alpha,\beta}(\nu)dc}{\pi_0\int_0^{\pi_1}\nu u_{\alpha,\beta}(\nu)dc+\pi_1\int_{\pi_1}^1(1-\nu)u_{\alpha,\beta}(\nu)dc},
\end{equation}
where $\mu=\nu_{0}+\nu_{1}$, $T(\nu)$ is the optimal decision threshold that minimizes the cost $T$ under $\nu$, $\pi_{k}$ is the prior probability of the $k$th class of the data, and $u_{\alpha,\beta}(\nu)$ is a Beta distribution with parameters $\alpha$ and $\beta$. $L_{\alpha,\beta}$ represents the general loss under $u_{\alpha,\beta}$, and $L_{Max}$ represents the maximum loss. The value of H-measure ranges from 0 to 1, with higher values indicating better algorithm performance.

AUC-mROC \cite{muscoloni2022} applies the idea of NDCG to optimize AUC by transforming both axes of the ROC curve using logarithmic transformations. The transformed curve is called mROC curve, which consists of the points $(nmFPR@k, mTPR@k)$, where $k \in [1,|E^{P}|]$, and AUC-mROC is the area under the mROC curve. The mathematical formulae for $nmTPR$ and $nmFPR$ are:
\begin{equation*}
	nmTPR@k=\frac{\ln (1+TP@k)}{\ln (1+|E^{P}|)},
\end{equation*}
\begin{equation*}
	nmFPR@k=\frac{\ln (1+FP@k)}{\ln (1+|U-E|)}.
\end{equation*}
To normalize and realign the mROC curve of a random predictor to the diagonal, an additional normalization is applied to the transformed $TPR$, as:
\begin{equation}
	mTPR@k=nmFPR@k+\frac{nmTPR@k-\ln_{(1+|E^{P}|)} (1+FP@k \cdot \frac{|E^{P}|}{|U-E|})}{1-\ln_{(1+|E^{P}|)} (1+FP@k \cdot \frac{|E^{P}|}{|U-E|})} \cdot (1-nmFPR@k).
\end{equation}

\subsection{Grey Correlation Coefficient}
In this study, we utilize the grey correlation coefficient to measure the correlation between any two sequences of discriminabilities. Assuming that the two sequences to be analyzed are $X_{i}=(x_{i1},x_{i2},\dots,x_{im})$ and $X_{j}=(x_{j1},x_{j2},\dots,x_{jm})$, where $m=8$ in this study. The correlation of sequence $X_{i}$ to sequence $X_{j}$ is given by:
\begin{equation}
	\xi_{ij}=\frac{1}{m}\sum_{e=1}^{m}\frac{\min_{i}\min_{t}|x_{jt}-x_{it}|+\rho \max_{i}\max_{t}|x_{jt}-x_{it}|}{|x_{je}-x_{ie}|+\rho \max_{i}\max_{t}|x_{jt}-x_{it}|},
\end{equation}
where $\rho \in [0,1]$ is a free parameter and set to be 0.5 in this study, $\min_{i}\min_{t}|x_{jt}-x_{it}|$ is the minimum difference, and $\max_{i}\max_{t}|x_{jt}- x_{it}|$ is the maximum difference \cite{deng1982}. From the above equation, the correlation of sequence $X_{j}$ to sequence $X_{i}$ can analogously be obtained as $\xi_{ji}$. Note that, in general,  $\xi_{ij} \neq \xi_{ji}$. The grey correlation coefficient between sequence $X_{i}$ and sequence $X_{j}$ is
\begin{equation}
	R_{ij}=\frac{\xi_{ij}+\xi_{ji}}{2}.
\end{equation}

If we would like to measure the correlation between discriminability sequences of the 8 metrics for different link prediction algorithms, then $X_{i}$ and $X_{j}$ represent the discriminability sequences for the $i$th and $j$th algorithm, while if we aim to measure the correlation between network domains, then $X_{i}$ and $X_{j}$ could be the discriminability sequences the $i$th and $j$th domains.

\bibliographystyle{plain}	

\end{document}